\documentclass[useAMS,usenatbib]{mn2e}

\usepackage{times}
\usepackage{bm}% bold math
\usepackage{graphicx}
\usepackage{dcolumn}%
\begin{document}

\journal{astro-ph/0603833}

\title[The Impact of Cosmic Dust on Supernova Cosmology]
{The Impact of Cosmic Dust on Supernova Cosmology}
\author[P. S. Corasaniti] {Pier Stefano Corasaniti\thanks{email: pierste@astro.columbia.edu} \\ 
{\em ISCAP, Columbia University, New York, NY 10027, USA} \\
{\em Department of Astronomy, Columbia University, New York, NY 10027, USA}
}
\maketitle

\begin{abstract}
Extinction by intergalactic gray dust introduces a magnitude redshift dependent offset 
in the standard-candle relation of SN Ia. This leads to overestimated luminosity distances 
compared to a dust-free universe. Quantifying the amplitude of 
this systematic effect is crucial for an accurate determination of the dark energy
parameters. In this paper we model the gray dust extinction 
in terms of the star-formation history of the Universe and the physical properties of the dust grains.
We focus on a class of cosmic dust models which satisfy current observational constraints. These can
produce an extinction as large as $0.08 \,{\rm mag}$ at $z=1.7$ and potentially disrupt the dark energy parameter
inference from future SN surveys. In particular depending on the dust model we find that
an unaccounted extinction can bias the estimation of a constant dark energy equation of state $w$
by shifting its best fit value up to $20\%$ from its true value.
Near-IR broadband photometry will hardly detect this effect, while the induced decrement of the Balmer lines
requires high signal-to-noise spectra. Indeed IR-spectroscopy will be needed for high redshift
SNe. Cosmic dust extinction may also cause a detectable violation of the distance-duality relation.
A more comprehensive knowledge of the physics of the IGM is necessary for an accurate
modeling of intergalactic dust. Due to the large magnitude dispersion current luminosity distance
measurements are insensitive to such possible extinction effects. In contrast these must be taken into account
if we hope to disclose the true nature of dark energy with the upcoming generation of SN Ia surveys.
\end{abstract}
\begin{keywords}
cosmology: theory - dust, extinction - supernova Ia - dark energy
\end{keywords}

%%%%%%%%%%%%%%%%%%%%%%%%%%%%%%%%%%%%%%%%%%%%%%%%%%%%%%%%%%%%%%%%%%%%%%

\section{Introduction}

Dust particles are present in the interstellar medium
causing the absorption of nearly $30-50\%$ of light emitted by stars
in the Galaxy. On the other hand very little is known about
dust particles which may exists outside our galactic environment. 
Metal lines are observed in the X-ray spectra of galaxy clusters
(e.g. Buote 2002) and in high redshift Lyman-$\alpha$ clouds (Cowie et al. 1995; Telfer et al. 2002). 
Infrared emissions of distant quasars have been attributed to the presence of large amounts
of dust (Bertoldi et al. 2003; Robson et al. 2004). Therefore it has been
speculated that some type of dust may be present in the low density
intergalactic medium (IGM). Since conditions in the IGM are unfavorable
to the formation of dust grains, if such a component exists it 
originates in stars. However it is unlikely that its properties
are similar to those of interstellar grains. In fact because of the physical processes
which expel dust from formation sites, intergalactic
dust particles may undergo very different selection effects
(Shustov \& Vibe 1995; Davies et al. 1998; Aguirre 1999).

Since the early search for distant Supernovae Type Ia (SNe Ia) 
(Riess et al. 1998; Perlmutter et al. 1999), cosmic dust extinction 
was proposed to account for the observed dimming of supernova
luminosities at high redshift (Aguirre 1999). 
From several other observations we have now compelling evidence 
of the cosmological nature of this signal 
(De Bernardis et al. 2000; Percival et al. 2001; Spergel et al. 2003; Scranton et al. 2003; Tegmark et al. 2004).
There is a general consensus that it
is caused by a recent accelerated phase of expansion
driven by a dark energy component. 
This can be the manifestation of a cosmological constant, or an exotic specie of matter, 
or a different regime of gravity on the large scales. 
Distinguishing between these
different scenarios has motivated a rich field of investigation.

Over the next decade numerous experiments will test dark energy 
using a variety of techniques.
Surveys of SN Ia such as the proposed {\em SNAP}, {\em JEDI} or ALPACA will play a leading role
by providing very accurate luminosity distance measurements. 
The success of this program will mostly depend 
on the ability to identify and reduce possible sources of
systematic uncertainties affecting the SN Ia standard-candle relation.

Here we address the impact of cosmic ``gray''
dust. Our aim is to study this particular systematic effect
from an astrophysical point of view. Differently from the original
proposal by Aguirre (1999), we do not look for dust models which
mimic the dimming of an accelerating universe. Instead 
we estimate how an hypothetical
cosmic gray dust model which satisfies existing astrophysical constraints
may affect the dark energy parameter estimation from future SN observations.
In order to do so we evaluate the dust extinction from first principles
by modeling the IGM dust in terms of the
star-formation history (SFH) of the Universe and the physical properties of the
dust grains. This will allow us to establish how the uncertainties
in the cosmic dust model, which depends on the complex physics of the IGM,
relate to expected cosmological parameter errors.

The paper is organized as follows: in Section \ref{cde} we discuss
the existing constraints on cosmic gray dust and evaluate
the expected extinction for different IGM dust models.
We evaluate the impact on the dark energy parameter inference and
describe the results of our analysis in Section \ref{sm}. In Section \ref{nir}
we compute the signature of dust models in near-infrared photometric
measurements and the decrement of the Balmer lines. 
In Section~\ref{rr} we discuss the violation of the distance-duality
relation. Finally in Section~\ref{conclu} we present our conclusions.

\section{Cosmic Gray Dust}\label{cde}

\subsection{Observational Constraints}
Typical dust extinction is correlated with reddening of incoming light, therefore
it can be revealed by simple color analysis.
Using this technique the interstellar extinction law has been estimated
over a wide range of wavelengths (e.g. Cardelli et al. 1989).
However this method is not effective for absorption caused
by ``gray'' dust. As suggested by Aguirre (1999),
astrophysical processes which transfer dust
into the IGM can preferentially destroy small grains over the large ones.
Those surviving have radii $a\ga 0.01 \mu{m}$.
In such a case intergalactic dust may consist of particles which induce
very little reddening (hence gray), while still able to cause large extinction effects.

The possibility of gray dust being entirely responsible for the dimming of high
redshift SN Ia has been now ruled out. For instance Aguirre \& Haiman (2000)
showed that the density of dust necessary to reconcile SN data
with a flat matter dominated universe is incompatible
with the limits inferred from the far-infrared background (FIRB) as measured
by the DIRBE/FIRAS experiment. 
Recently Bassett and Kunz (2004b) have excluded this scenario
at more than $4\sigma$ by testing the distance-duality relation.
Nevertheless the actual amount of dust in the IGM and its composition remain unknown.

Constraints on cosmic dust extinction have been inferred from 
color analysis of distant quasars (Mortsell \& Goobar 2003; 
Ostman \& Mortsell 2005). Assuming the interstellar extinction 
law (Cardelli et al. 1989; Fitzpatrick 1999), these studies 
have confirmed that dust dimming cannot be larger than $0.2\,{\rm mag}$ at $z=1$
and also indicated that if any gray dust component is present in the
IGM it cannot induce extinction larger than $0.1\,{\rm mag}$.
For an early study of the effect of intergalactic extinction on cosmological expansion
measurements see also Meinel (1981).

Indeed infrared (IR) observations may turn out to be more informative.
As an example Aguirre \& Haiman (2000) have suggested that 
resolving the Far Infrared Background (FIRB) will provide a definitive test of the IGM dust.
Some quantitative limits have also been derived from the thermal
history of the IGM (Inoue \& Kamaya 2003). 

A more direct constraint on the density of cosmic dust particles has been obtained 
by Paerels et al. (2002) from the analysis of X-ray scattering halo around
a distant quasar at $z=4.30$. In particular for grains of size $\sim1\,\mu{m}$
the total cosmic dust density is $\Omega_{\rm dust}^{\rm IGM}\la 10^{-6}$, while for $0.1\,\mu{m}$ grains
the constraint is one order of magnitude weaker.
Compatible limits were also found by Inoue \& Kamaya (2004) using existing bounds
on SN Ia extinction and reddening. 

As we will see in the next section a better knowledge of these quantities
is necessary if we hope to measure the dark energy parameters with high accuracy.

\subsection{Intergalactic Dust Extinction}\label{igm}
In order to estimate the extinction from intergalactic gray dust we need
to determine the evolution of dust density in the IGM.
Since dust particles are made of metals, the first step is to evaluate
the evolution of the cosmic mean metallicity from the star formation
history of the Universe (Aguirre \& Haiman 2000). For simplicity we can assume that metals
are instantaneously ejected from newly formed stars.
In such a case the metal ejection rate per unit
comoving volume at redshift $z$ can be written as (Tinsley 1980):
\begin{equation}
\dot{\rho}_{Z}(z)=\dot{\rho}_{\rm SFR}(z)y_{Z},\label{zrate}
\end{equation}
where $\dot{\rho}_{\rm SFR}$ is the star formation rate
and $y_Z$ is the mean stellar yield, namely the average mass fraction 
of a star that is converted to metals. The value of $y_Z$
is slightly sensitive to the initial mass function (IMF) and may also change 
with redshift if the IMF varies with time. For simplicity we assume $y_Z$ to be constant. 

From Eq.~(\ref{zrate}) it follows that the mean cosmic metallicity is given by
(Inoue \& Kamaya 2004):
\begin{equation}
Z(z)=\frac{y_Z}{\Omega_{\rm b}\rho_c}\int^{z_S}_{z}\dot{\rho}_{\rm SFR}(z')\frac{dz'}{H(z')(1+z')},
\end{equation}
where $\Omega_b$ is the baryon density, $\rho_c$ is the current critical density, $H(z)$
is the Hubble rate and $z_S$ redshift at which star formation began.
There is little dependence on $z_S$ for $z\la3$, provided that the star formation
begin at $z_S\ga 5$. Without loss of generality we set its value to $z_S=10$.

Following the notation of Inoue \& Kamaya (2004), we introduce a further parameter
which describes the mass fraction of dust to the total metal mass,
$\chi=\mathcal{D}/Z$, where $\mathcal{D}$ is the dust-to-gas ratio of the IGM. 
The latter depends on the mechanism which expel dust from galaxies and in principle
may evolves with redshift according to the dominant process responsible for 
the transfer (e.g. stellar winds, SN II explosions, radiation pressure). 
Only recently authors 
have began to study the metal enrichment of the IGM using numerical simulations 
(see for instance Bianchi \& Ferrara 2006). As we lack of any guidance we simply assume 
that the dust-to-gas ratio scales with the mean metallicity and 
consider $\chi$ as a constant free parameter. 

Another open issue concerns the spatial distribution of dust particles in the IGM. 
It has been argued that a clumped gray dust would cause a dispersion of supernova magnitudes larger
than the observed one. Consequently if a gray dust component exists it must be nearly homogeneously 
distributed. However this does not necessarily implies a strong constraint on the
gray dust hypothesis. In fact the overall dispersion at a given redshift goes roughly
as $\Delta\propto 1/\sqrt{N}$ where $N$ is the number of homogeneous dust patches
along the line-of-sight (Aguirre 1999). 
Numerical simulations indicate that dust grains can diffuse in one billion years 
over scales of few hundreds Kpc (Aguirre et al. 2001). This
corresponds to $N\gg1$ for high redshift SNe, in which case the dispersion would be small.
Indeed more detailed studies are needed, here we limit our analysis 
to a homogeneous dust distribution.

The differential number density of dust particles
in a unit physical volume reads as 
\begin{equation}
 \frac{dn_{\rm d}}{da}(z)=\frac{\chi\, Z(z)\,\Omega_{\rm b}\rho_c(1+z)^3}
  {4\pi a^3 \varrho/3}N(a),\\
\end{equation}
where $\varrho$ is the grain material density and $N(a)$ is the grain size distribution normalized
to unity.

The amount of cosmic dust extinction on a source at redshift $z$ observed
at the rest-frame wavelength $\lambda$ integrated over the grain size distribution is
then given by:
\begin{equation}
 \frac{A_{\lambda}(z)}{\rm mag} =1.086 \pi \int_0^z \frac{c\, dz'}{(1+z')H(z')}
\int a^2 Q_m^{\lambda}(a,z') \frac{dn_{\rm d}}{da}(z')d{a},\,
\label{ext}
\end{equation}
where $Q_m^{\lambda}(a,z')$ is the extinction efficiency factor which depends on the grain size $a$ and complex
refractive index $m$, and $c$ is the speed of light. Hence the extinction at a given
redshift depends on the dust properties and the metal content of the IGM.
More specifically for a given cosmological background, a model of dust is specified
by the grain composition, size distribution and material density,
the mean interstellar yield, the star formation history and
the IGM dust-to-total-metal mass ratio.

None of these parameters is precisely known, leaving us
with a potentially large uncertainty about the level of cosmic dust extinction. 

In the following we assume a standard flat LCDM model with
Hubble constant $H_0=70 \,{\rm Km\,s^{-1}\,Mpc^{-1}}$, matter density $\Omega_{\rm m}=0.30$ and 
baryon density $\Omega_b=0.04$.

Several studies have suggested that the size of IGM dust grains
varies in the range $0.05-0.2\,\mu{m}$ (Ferrara et al. 1991; Shustov \& Vibe 1995; Davies et al. 1998). 
Smaller grains ($a\la 0.05 \mu{m}$) are either destroyed by sputtering or
unable to travel far from formation sites as they are inefficiently pushed away by radiation pressure;
in contrast grains larger than $\sim0.2 \mu{m}$ are too heavy and remain trapped in the gravitational 
field of the host galaxy. However these analysis have provided no statistical description 
of the grain size abundance. A common assumption is to consider a power law distribution, 
$N(a)\propto a^{-3.5}$, usually referred as the MRN model (Mathis, Rumpl \& Nordsiek 1977). 
This describes the size distribution of dust grains in the Milk Way, but there is no 
guarantee that this model remains valid for IGM dust as well. On the other hand Bianchi \& Ferrara (2005)
have studied through numerical simulation the size distribution of grains ejected into the IGM. 
Assuming an initial flat size abundance they find that the post-processed 
distribution remains nearly flat and due to erosion sputtering the size range is 
slightly shifted towards smaller radii, $0.02-0.15\,\mu{m}$.
We refer to this as the BF model and evaluate the gray dust extinction for both MRN and BF cases. 
We also consider a uniformly sized dust model corresponding to a 
distribution $N(a)= \delta(a)$, with 
$a=0.1\mu{m}$ and for more descriptive purpose we also consider the less realistic value $a=1.0\mu{m}$. 

The exact intergalactic dust composition is also not known, we focus 
silicate and graphite particles with material density $\varrho=2{\rm g\,cm^{-3}}$,
and optical properties specified as in Draine \& Lee (1984).
Using these specifications we compute the extinction efficiency factor $Q_m^{\lambda}(a,z)$
by solving numerically the Mie equations for spherical grains (Barber \& Hill 1990).

We set the mean interstellar yield to $y_Z=0.024$ (Madau et al. 1996) corresponding
to the value inferred from the Salpeter IMF (Salpeter 1955). 

\begin{figure*}
\includegraphics[width=8cm]{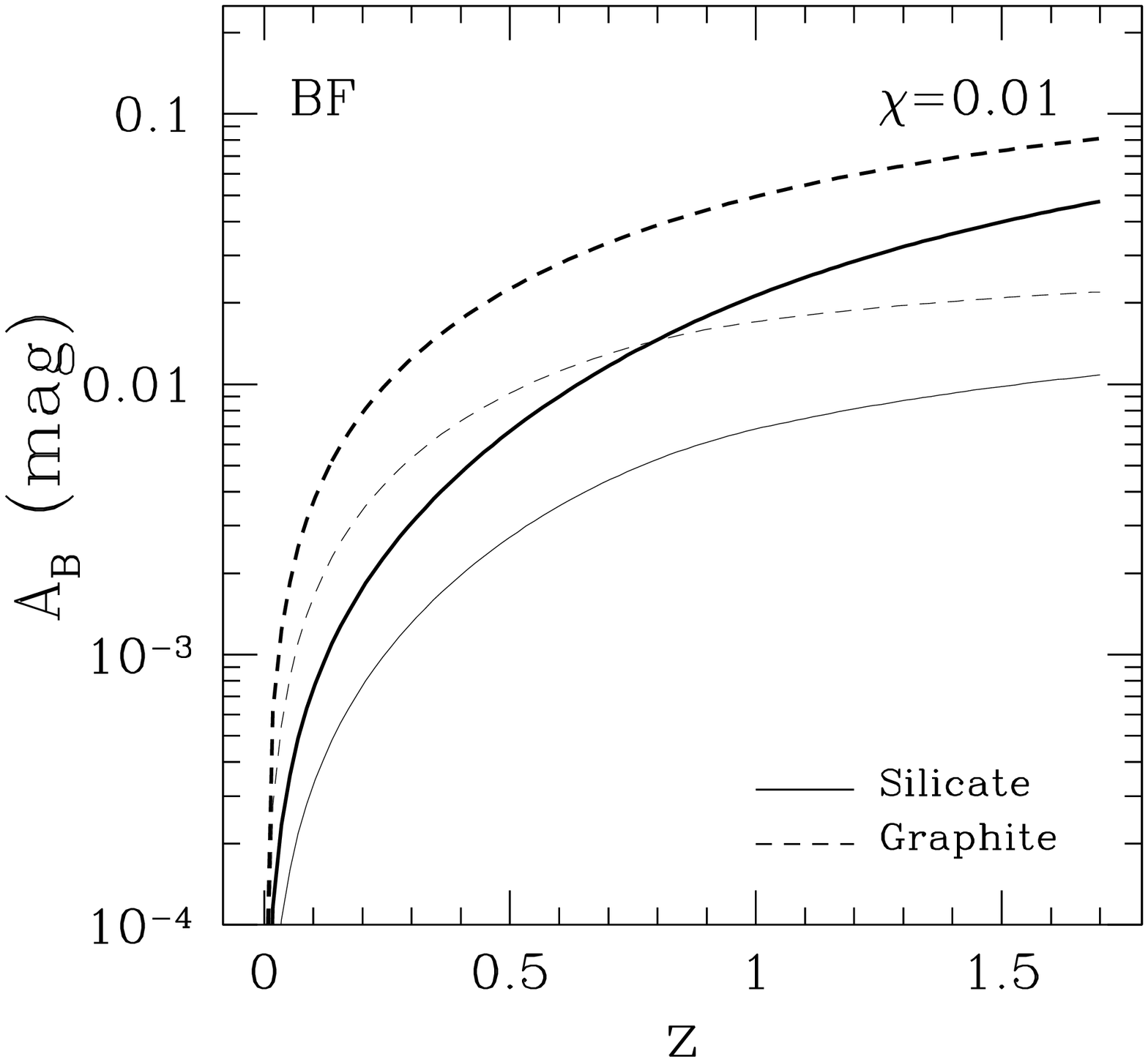}
\includegraphics[width=8cm]{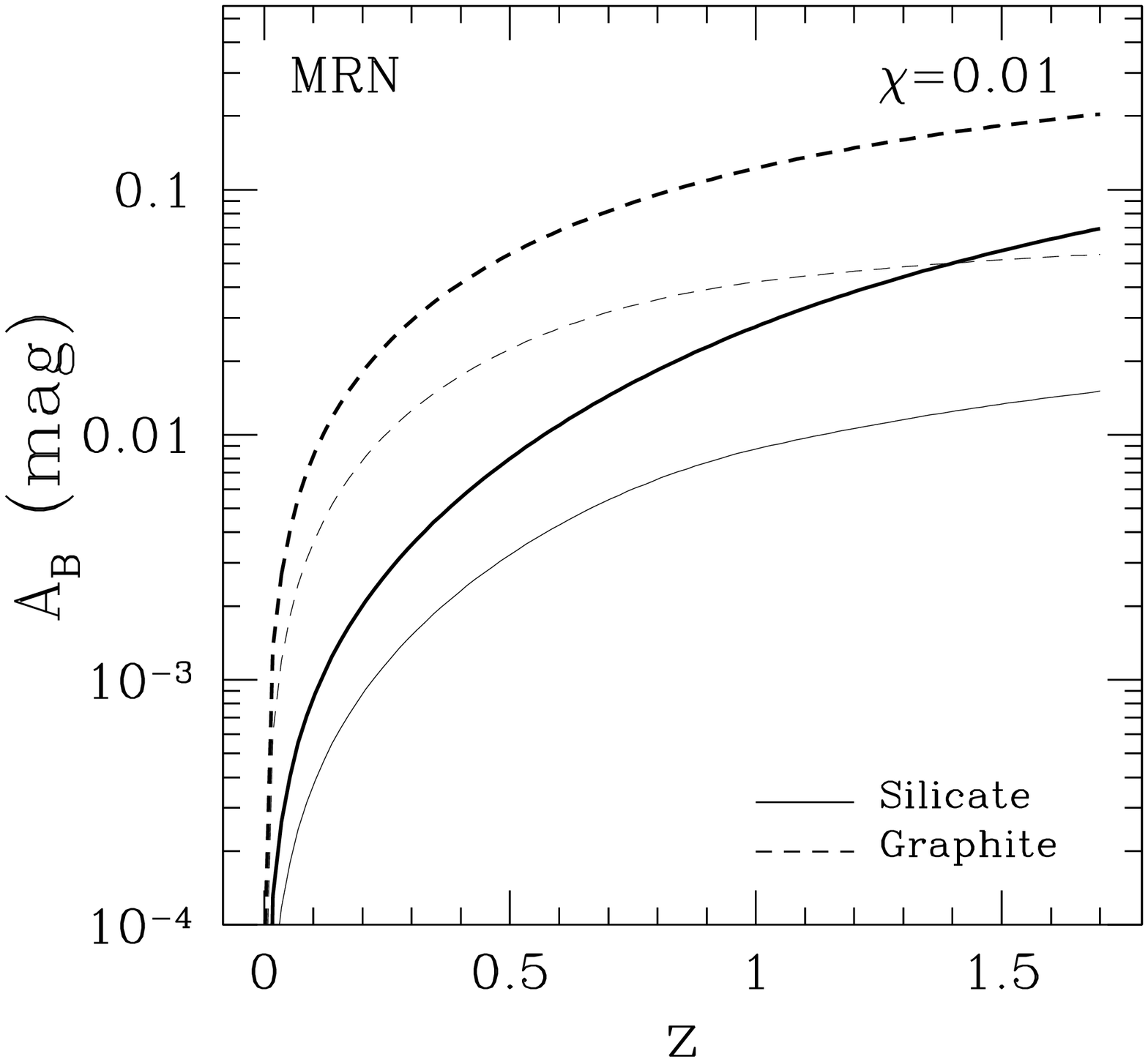}
\includegraphics[width=8cm]{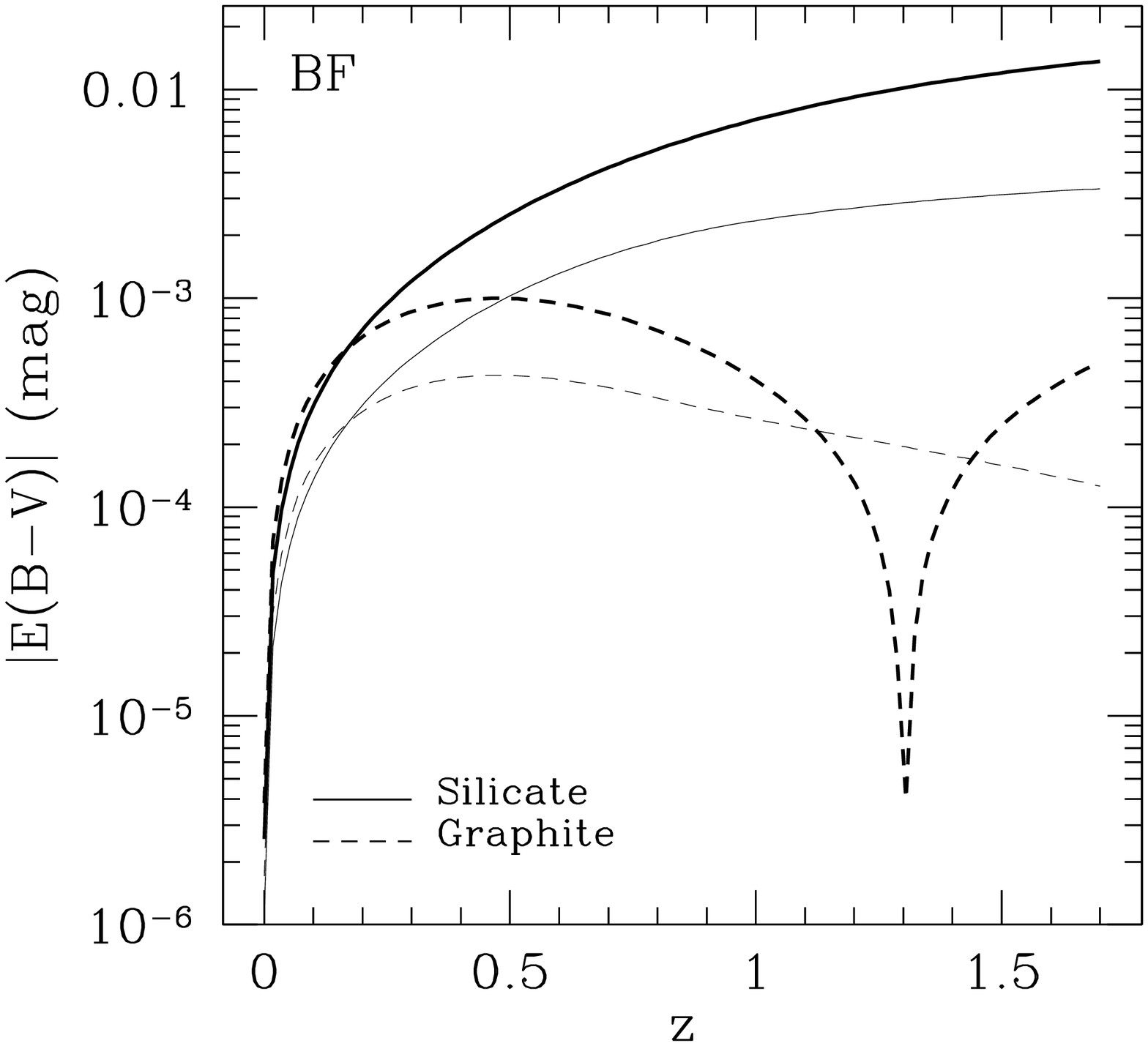}
\includegraphics[width=8cm]{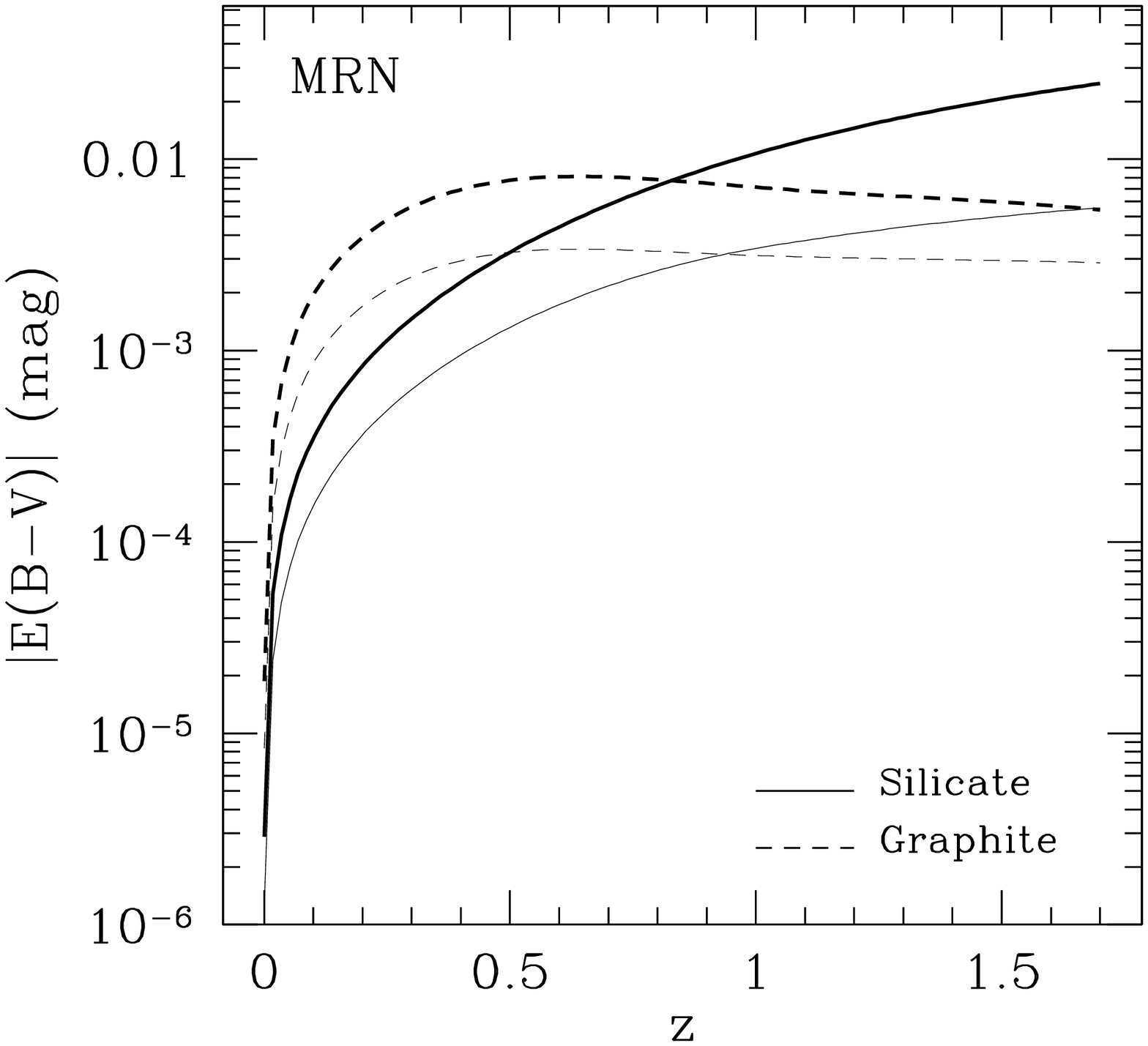}
\caption{Cosmic gray dust extinction in the B-band (upper panels) 
and color excess (lower panels) as function of redshift of the source
for BF (left panel) and MRN (right panel) grain size distributions in the range $0.02-0.15\,\mu{m}$. 
Solid and dash lines correspond to silicate and graphite grains
respectively. Thick (thin) lines correspond to high (low) SFH models.
}
\label{fig1}
\end{figure*}
\begin{figure*}
\includegraphics[width=8cm]{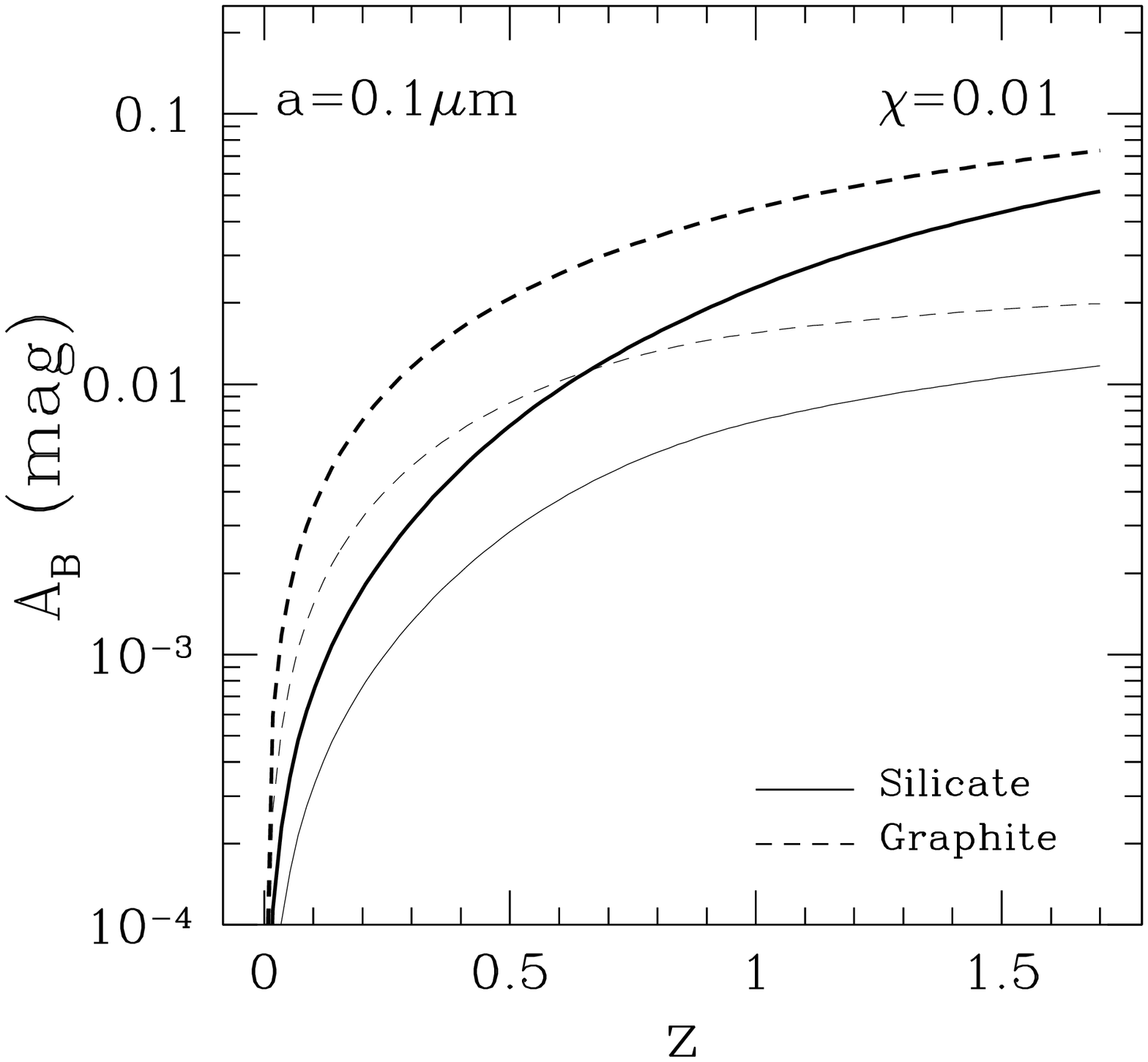}
\includegraphics[width=8cm]{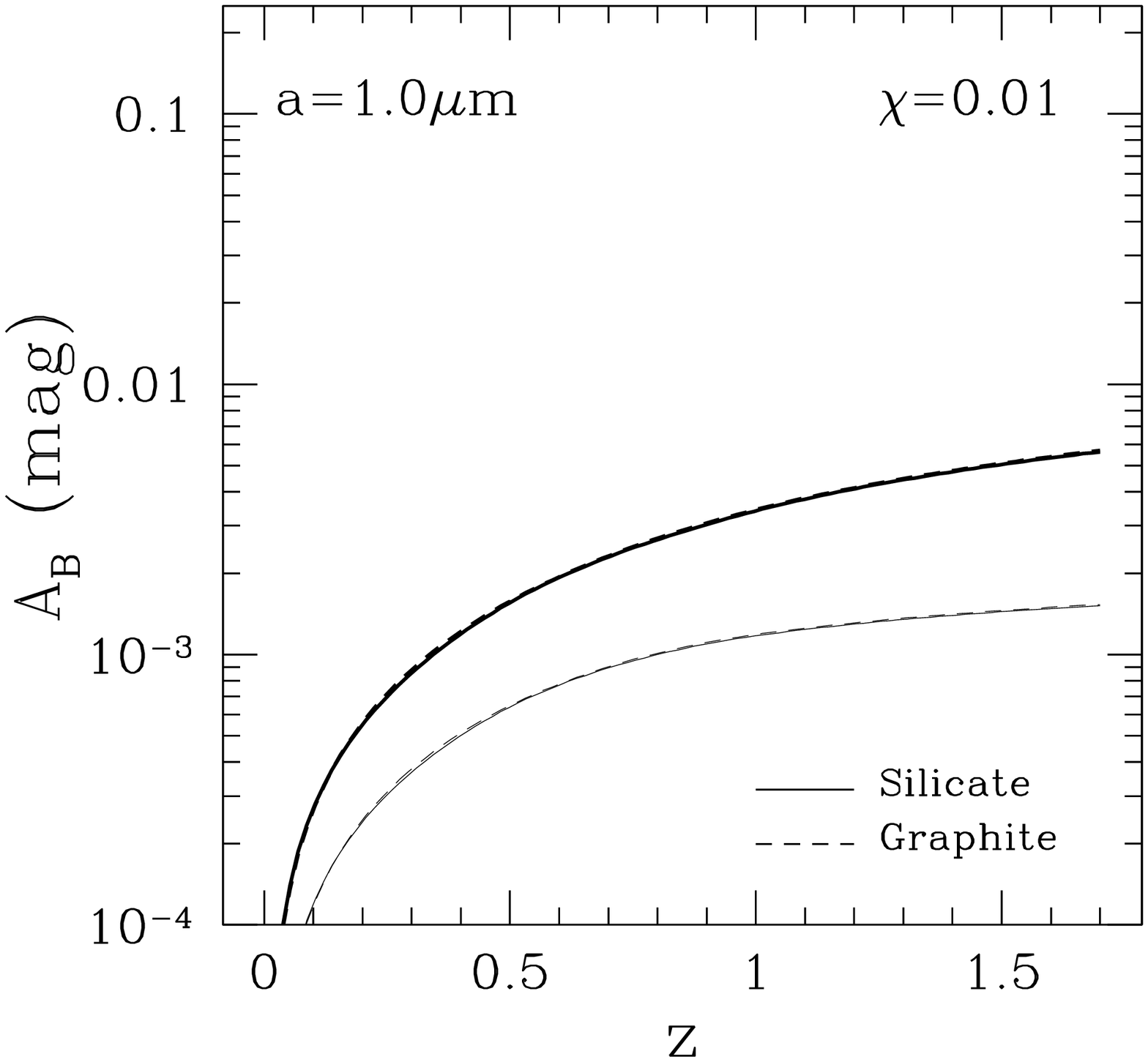}
\includegraphics[width=8cm]{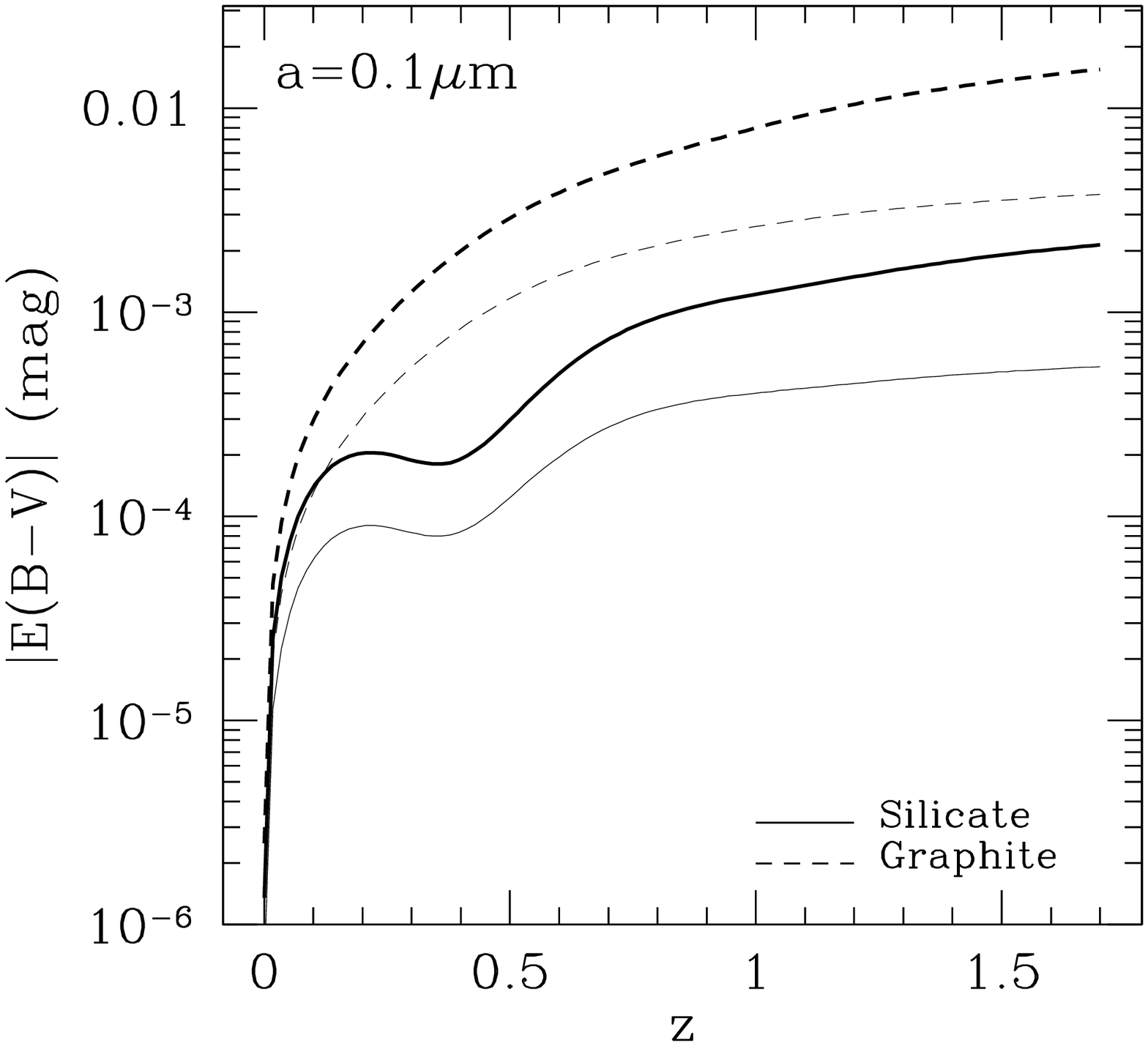}
\includegraphics[width=8cm]{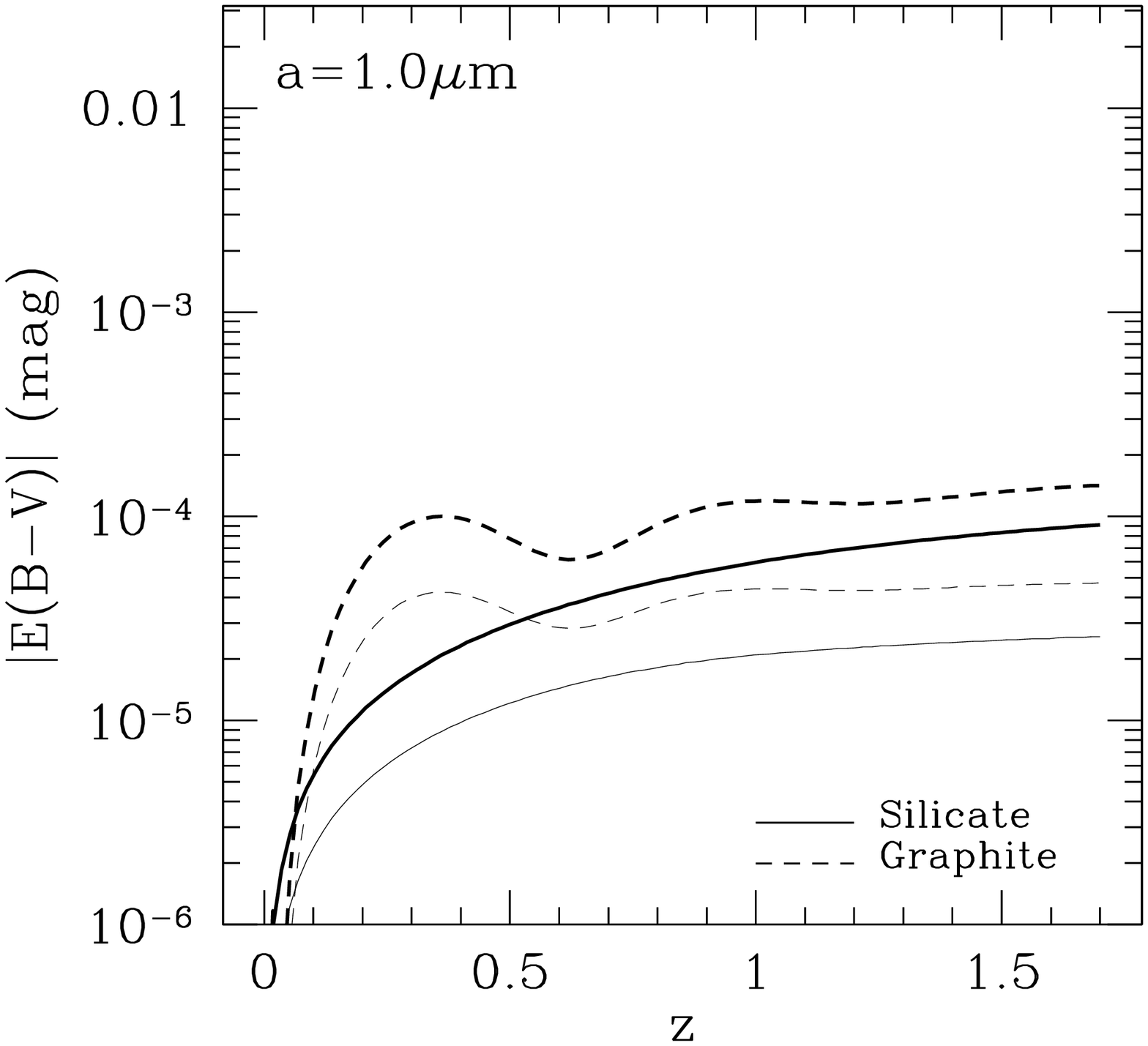}
\caption{As in Fig.~\ref{fig1}
for a uniform sized grains with $a=0.1\,\mu{m}$ (left panel) and $a=1.0\,\mu{m}$ (right panel). 
}
\label{fig2}
\end{figure*}
The star formation rate at different redshifts is known from a large body of measurements.
The trend at redshifts $z\le 1$ is well established with
\begin{equation}
\frac{\dot{\rho}_{SFR}(z)}{\rm M_{\odot}\, yr^{-1} Mpc^{-3}}=0.0158\, (1+z)^{3.10},
\end{equation} 
being the best fit to existing data (Hopkins 2004). 
On the other hand there is less agreement on the exact behavior at higher
redshifts, with recent observations favoring a flat redshift dependence (Giavalisco et al. 2004). 
We follow the analysis of Inoue and Kamaya (2004) 
and consider two possible star-formation rates at $z>1$:
\begin{equation}
\frac{\dot{\rho}_{SFR}(z)}{\rm M_{\odot}\, yr^{-1} Mpc^{-3}}=\cases{
0.136 & (high SFH) \cr
0.384\,(1+z)^{-1.5} & (low \,SFH) \cr
}
\end{equation}
in units of solar mass $M_{\odot}$ per year per Mpc volume.

Consistently with constraints derived in (Inoue \& Kamaya 2004) we set $\chi=0.01$.
Since Eq.~(\ref{ext}) is linear in this parameter the results can be simply 
rescaled for different values. For this particular choice
the total dust density up to $z=4.3$ is $\Omega_{\rm dust}^{\rm IGM}\sim 10^{-6}$,
which is consistent with the direct constraints found in Paerels et al. (2002).
In addition dust grains in the IGM can absorb the UV light in the Universe and re-emit in the far infrared
contributing the FIRB. From the analysis of Aguirre \& Haiman (2000) we find that for $\chi=0.01$
cosmic gray dust would produce a background signal at $850\,\mu{m}$ roughly $10\%$ of the 
FIRB and only $1\%$ at $200\,\mu{m}$, thus well within the DIRBE/FIRAS limits.

In figure \ref{fig1} we plot the B-band extinction (upper panels) and
reddening (lower panels) as function of the redshift for BF (left panels) and MRN (right panels)
grain size distributions.
The solid and dash lines correspond to silicate and graphite grains 
respectively. Thick (thin) lines correspond to high (low) SFH models.
Low SFH gives smaller extinction than the high case, 
consistently with the fact that low SFH produce a smaller amount of dust.
The extinction is larger for graphite grains than silicate. Notice also 
that the extinction for the BF distribution is smaller than for the MRN case.
This is because in the B-band the efficiency factor is constant,
thus Eq.~(\ref{ext}) scales as $N(a)/a$. Since smaller grains are more
abundant in the MRN model than in the BF case, the corresponding extinction is larger.

As it can be seen from the plots of the color excess $|E(B-V)|$ these
models cause very little reddening. Photometric measurements 
more accurate than $1\%$ would be needed to detect the imprint of gray dust at high redshift. 

In figure \ref{fig2} we plot the case of uniformly sized grains with radii $a=0.1\,\mu{m}$
and $a=1.0\,\mu{m}$. As expected $a=0.1\,\mu{m}$ grains cause an extinction nearly a factor $10$
larger than $1.0\,\mu{m}$ particles, consistently with the $1/a$ dependence of $A_B$. 
Although these models are unrealistic from a purely astrophysical stand point, we can see that
for $a=0.1\mu{m}$ the expected extinction and reddening are in agreement with 
those estimated assuming more realistic grain size distributions. Therefore without loss of generality 
we can use the uniform size approximation to study 
the effect of dust extinction on the dark energy parameter
inference without the need to specify the exact form of $N(a)$. We can simply
focusing on the typical size of gray particles and the other parameters specifying the IGM dust model.

\section{Dark Energy Inference}\label{sm}

Supernova Type Ia observations measure the luminosity distance
through the standard-candle relation,
\begin{equation}
m_{B}(z)=\mathcal{M}_{B}+5\log{H_0\,d_L(z)},\label{mz}
\end{equation}
where $m_{B}(z)$ is the apparent SN magnitude in the B-band,
$\mathcal{M}_{B}\equiv M_{B}-5\log{H_0}+25$ is the ``Hubble-constant-free''
absolute magnitude and $d_L(z)$ is the luminosity distance.

Extinction modifies the standard-candle relation 
such that the observed SN magnitude is
\begin{equation}
\tilde{m}_{B}(z)=m_{B}(z)+A_{B}(z),\label{mdust}
\end{equation}
with $A_{B}(z)$ given by Eq.~(\ref{ext}) evaluated at the B-band center rest-frame
wavelength, $\lambda=0.44\,\mu{m}$. Hence
supernovae are systematically dimmer than in a dust-free universe,
and overestimate luminosity distances. 
Note that the extinction term in Eq.~(\ref{mdust})
corresponds to a redshift dependent magnitude offset. 
Previous studies of supernova systematics have limited their analysis to
a simple magnitude offset that linearly increases with redshift 
(Weller \& Albrecht 2002; Kim et al. 2004). On the contrary here we approach this
type of systematic from a physically motivated standpoint.
Having modeled the gray dust extinction as in Eq.~(\ref{ext}), 
we can determine how astrophysical uncertainties in the cosmic dust
model parameters affect dark energy parameter inference. 

\subsection{Monte Carlo Simulations}
Using Eq.~(\ref{mdust}) we proceed by Monte Carlo 
simulating a sample of SN Ia data in the B-band in a given cosmological background
for dust models listed in table~\ref{tab1}. 
Then for each of these samples we recover the background cosmology by
inferring the best fit dark energy parameter values 
and uncertainties in a dust-free universe through standard likelihood analysis.

We consider
a constant dark energy equation of state $w$ and 
a time-varying equation of state of the form
(Chevallier \& Polarski 2001; Linder 2003)
\begin{equation}
w(z)=w_0+w_1 \frac{z}{1+z}.\label{lind}
\end{equation} 
For simplicity we focus on a SN experiment such as SNAP which goes very far in redshift.
We assume the survey characteristics as specified in (Kim et al. 2004). 
We consider a flat universe with $\Omega_{\rm m}=0.3$ 
and assume a Gaussian matter density prior $\sigma_{\Omega_{\rm m}}=0.01$. 

\begin{table}\centering\caption{Grey dust models. For $a=1.0\,\mu{m}$ we only 
consider Silicate dust since Graphite causes the same extinction.}\label{tab1}
\begin{tabular}{ccccc}
\hline
&$\chi$ & a & Type & SFH \\
\hline
A&0.01 & 0.1 & Graphite & low/high \\
B&0.01 & 0.1 & Silicate & low/high \\
C&0.01 & 1.0 & Silicate & low/high \\
\end{tabular}
\end{table}

\begin{figure}
\includegraphics[width=8cm]{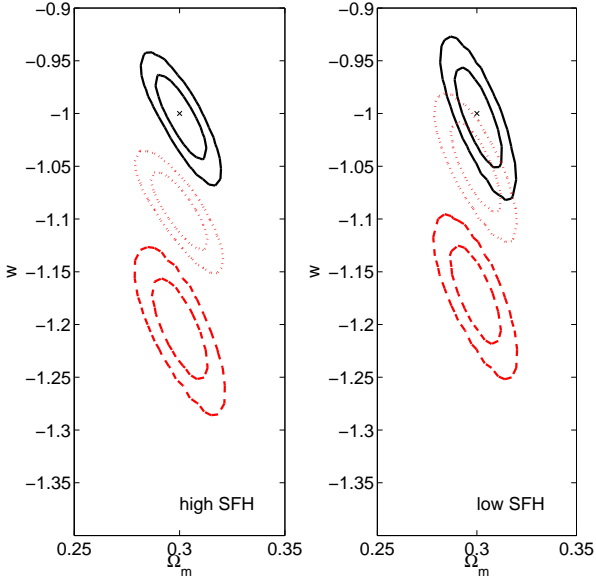}
\caption{Marginalized $1$ and $2\sigma$ confidence contours in the plane $\Omega_m-w$ plane
inferred from data generated in models A (red dash), B (red dot) and C (black solid) in LCDM background. The left and right panels
correspond to high and low SFH models respectively. The cross point indicates the parameter values
of the fiducial cosmology.}\label{fig3}
\end{figure}

First we consider the case of a fiducial LCDM cosmology. In figure~\ref{fig3} we plot 
the marginalized $1$ and $2\sigma$ contours in the $\Omega_{\rm m}-w$ plane
inferred from the data samples generated in models A (red dash), 
B (red dot) and C (black solid) for
low (left panel) and high SFH (right panel). 
It can be seen that the overall effect of extinction is to shift the confidence regions
towards more negative value of the dark energy equation of state.
This is because the extinction dims supernovae increasingly with
the redshift. Thus inferred distances are bigger than in a dust-free universe
mimicking a more rapid accelerating expansion. For fixed values of $\Omega_{\rm m}$ this requires
the dark energy equation of state to be $<-1$.
As a result an unaccounted extinction moves the best fit dark energy model 
many sigma away from the true one. The effect
is more dramatic in model A since $A_B(z)\ga0.01$ at $z>0.5$, while 
it is negligible in model C since the extinction is a factor ten smaller.
From figure~\ref{fig3} it is evident that the existence of
gray dust particles with size $\sim0.1\mu{m}$
and a dust-to-total metal mass ratio of $0.01$ in LCDM cosmology would
cause an extinction that effectively mimic a phantom dark energy model, hence
misleading us on the true nature of dark energy.

\begin{figure}
\includegraphics[width=8cm]{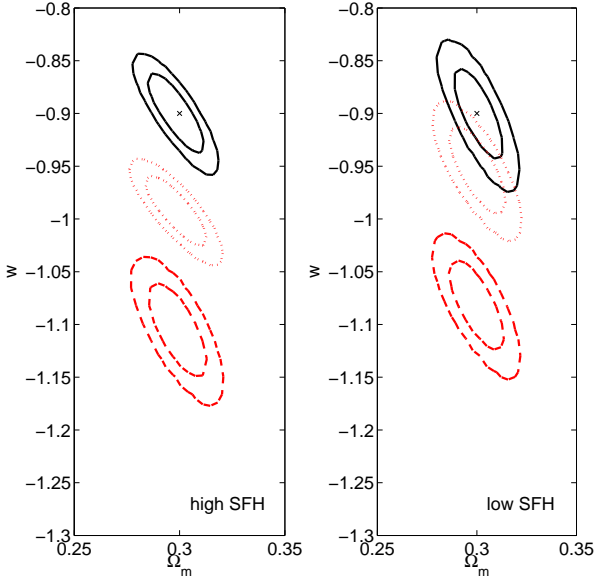}
\caption{As in figure~\ref{fig3} with $w=-0.9$ dark energy fiducial cosmology.}\label{fig4}
\end{figure}

In the same manner IGM dust may prevent us from detecting a quintessence-like
dark energy. For instance in figure~\ref{fig4} we plot the confidence contours in the 
case of a fiducial dark energy cosmology with $w=-0.9$. 
Again the effect of dust extinction is to shift the confidence regions towards 
more negative values of $w$. The amplitude of this effect is similar to the previous LCDM case 
and therefore is fiducial cosmology independent.

To be quantitative the extinction in model A causes a $20\%$ bias on the 
inferred values of $w$ and $10\%$ in model B. On the contrary model C
does not affect the parameter inference.

\begin{figure}
\includegraphics[width=8cm,clip]{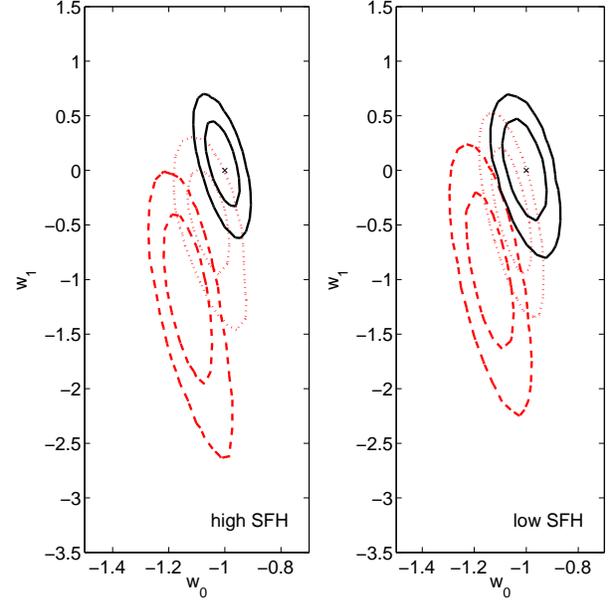}
\caption{Marginalized $1$ and $2\sigma$ contours in the plane $w_0-w_1$ for
dust models as in figure~\ref{fig3}.}\label{fig5}
\end{figure}

A similar trend occurs for the constraints on the redshift parametrization Eq.~(\ref{lind}).
We plot in figure~\ref{fig5} the marginalized $1$ and $2\sigma$ contours in the $w_0-w_1$ plane
for a fiducial LCDM cosmology. Notice that the size of the ellipses
is altered, besides the amplitude of the shift is smaller
 than for the constant equation of state parameter. In fact while in model A
the fiducial cosmology still lies many sigma away from the $95\%$ confidence
region, it is within the $2\sigma$ contours for model B.
This is because the effect of the extinction is spread over two degenerate equation 
of state parameters. Indeed IGM dust parameters should be included in the cosmological 
fit along the line suggested by Kim \& Miquel (2006).

\subsection{SN-Gold Data Analysis}
Can dust extinction affect the dark energy parameters inference 
from current SN Ia data? Despite the recent progress in the search for SN Ia, 
the magnitude dispersion is still large ($\sim0.1\,{\rm mag}$). Therefore the extinction
effect is well within the experimental errors. As an example we consider
the Gold sample (Riess et al. 2004) which extends up to $z_{max}\sim1.7$ and therefore
is more likely to be sensitive to gray dust extinction than the SNLS dataset (Astier et al. 2006)
for which $z_{max}\sim1$. In addition the estimated SN extinctions in the Gold dataset appear
to be correlated with the magnitude dispersion (Jain \& Ralston 2006).
We assume a flat universe with prior $\Omega_{\rm m}=0.27\pm0.04$.
Using Eq.~(\ref{mdust}) we fit the Gold data accounting for the extinction of 
dust model A. We find $w=-0.90\pm^{0.17}_{0.21}$ at $1\sigma$. 
On the contrary the fit without extinction gives $w=-0.96\pm^{0.18}_{0.16}$. Thus the shift 
is less than $1\sigma$. The fact that the best fit value is slightly $>-1$ should not
be surprising. Comparison with the SNLS data shows that SNe in the Gold sample are slightly brighter.
Nevertheless the LCDM is within $1\sigma$ uncertainty. Notice that the direction of 
the shift is consistent with the result of the Monte Carlo analysis. 
In fact accounting for the extinction term allows models with a larger value of $w$ to be consistent
with the data. 

\section{Near-IR Color Analysis and Decrement of Balmer Lines}\label{nir}
\begin{figure}
\includegraphics[width=8cm]{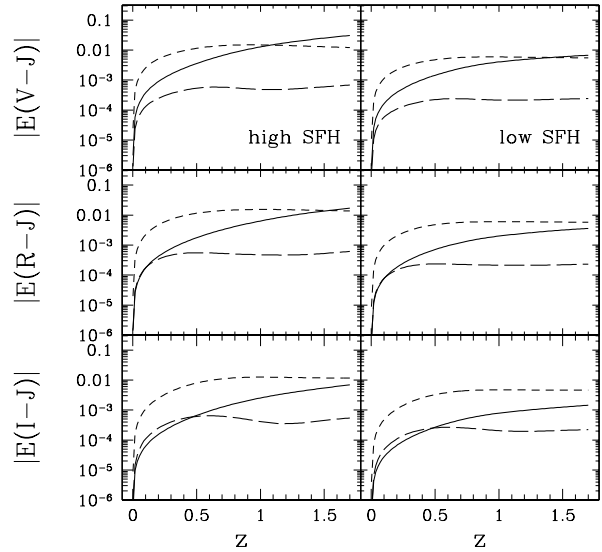}
\caption{Absolute value of color excess $E(V-J)$, $E(R-J)$ and $E(I-J)$ versus
redshift for models A (short-dash), B (solid) and C (long-dash) in the case of
high (left panel) and low SFH (right panel).}\label{fig6}
\end{figure}

As we have seen in Section~\ref{igm} it is very difficult to detect 
the signature of gray dust through reddening analysis in the 
optical wavelengths. It has been suggested that broadband 
photometry in the near-IR could be more effective. 
For instance Goobar et al. (2002) estimate in $1\%$
the spectro-photometric accuracy necessary to detect the dust reddening 
in the I, J and R bands.
In figure~\ref{fig6} we plot the color excess $|E(V-J)|$, $|E(R-J)|$ and $|E(I-J)|$
for our test-bed of cosmic dust models. For low SFH models the color
excess is to small too be detectable with $1\%$ photometry. Only model A
in the high SFH case would be marginally distinguishable. In general we
find that our estimates are a factor two smaller than in Goobar et al. (2002).
Given the difficulty of performing such accurate near-IR measurements,
distinguishing the effect of cosmic dust will be a challenging task. 

\begin{figure}
\includegraphics[width=8cm]{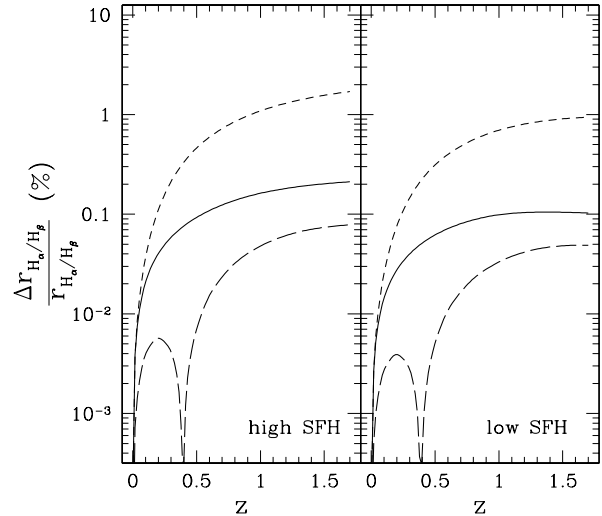}
\caption{Relative decrement of Balmer lines for dust models as in figure~\ref{fig6}.}\label{fig7}
\end{figure}

A possible alternative is to consider the decrement in the relative
strength of the Balmer lines in the host galaxy spectrum.
The recombination of ionized hydrogen atoms causes the well known
$H_\alpha$ and $H_\beta$ emission lines at $6563$ \AA \,and $4861$ \AA \,respectively
with intensity ratio $r_{H_\alpha/H_\beta}=2.86$. 
Deviations from this value are indicative of selective absorption. For instance
in the case of an extinction law with negative slope, the blue light is dimmed more
than the red one, hence causing $r_{H_\alpha/H_\beta}>2.86$. In figure~\ref{fig7} we plot
the absolute value of the relative decrement of the Balmer lines as function of 
redshift for models A, B and C respectively. We may notice that the amplitude of the decrement
for model A and B is within standard accuracy of high signal-to-noise spectroscopy. 
It is also worth noticing that for $1.0\,\mu{m}$ Silicate grains (model C) the
extinction law at $z>0.4$ changes slope causing $r_{H_\alpha/H_\beta}<2.86$.   
Deep redshift spectroscopic surveys can in principle be used to track the trend
of the Balmer line decrement and provide a complementary method to test the cosmic dust extinction.
However IR observations are necessary in order to measure the $H_\alpha$ emission of high redshift sources. 
As an example the SDSS catalog of galaxy and quasar spectra spans the range $3800<\lambda<9200$ \AA\,
therefore the $H_\alpha$ cannot be detected for objects at $z\ga 0.25$. 
The next generation of satellite surveyors 
will be equipped for IR-spectroscopy and capable to provide such measurements.

\section{Testing Distance-Duality Relation}\label{rr}

A well known result of metric theories of gravity is the uniqueness of
cosmological distances (Etherington 1933). Thus measurements of the luminosity
distance $d_L(z)$ and angular diameter distance $d_A(z)$ at redshift $z$ are
linked through the duality relation (Linder 1988; Schneider et al. 1992)
\begin{equation}
Y\equiv\frac{d_L(z)}{d_A(z)(1+z)^2}=1.\label{dual}
\end{equation}
As discussed in (Bassett \& Kunz 2004b) testing this equality with high accuracy
can be a powerful probe of exotic physics. Violations of the duality relation
are predicted by non-metric theories of gravity, varying fundamental constants and axion-photon
mixing (Bassett \& Kunz 2004a; Uzan et al. 2004) just to mention a few. Also astrophysical mechanisms 
such as gravitational lensing and dust extinction can cause deviation from Eq.~(\ref{dual}).

From Eq.~(\ref{mz}) and Eq.~(\ref{mdust}) it is easy to show 
that in the presence of dust extinction the deviation from Eq.~(\ref{dual})
is given by
\begin{equation}
\Delta{Y}(z)=10^{\frac{1}{5}A_{B}(z)}-1.\label{dy}
\end{equation} 
Therefore if SN Ia are dimmed by intergalactic dust absorption, this would be manifest
in the violation of the duality relation. 

The distance-duality relation can be tested using SN Ia data and angular diameter
distance measurements from detection of baryon acoustic oscillations (BAO)
in the galaxy power spectrum (Bassett \& Kunz 2004b; Linder 2005). Over the next decade several surveys of the large scale structures
will measure $d_A(z)$ with few percent accuracy over a wide range of redshifts.
Similarly future SN Ia surveys such as SNAP are designed to control intrinsic supernova 
systematics within few percent which would provide luminosity
distances measurements with $1-2\%$ accuracy. 

We forecast the sensitivity 
of future distance-duality test by error propagation of Eq.~(\ref{dual}).
We assume the expected errors on the angular diameter distance
for a galaxy survey of $10,000\,{\rm deg^2}$ with 
spectroscopic redshifts as quoted in (Glazebrook \& Blake 2005). 

\begin{figure}
\includegraphics[width=8cm]{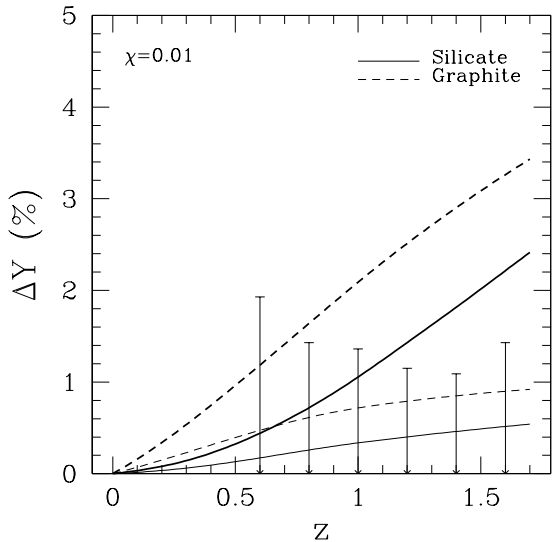}
\caption{Percentage deviation from the distance-duality relation as function
of redshift. The dash and solid lines represent the violation caused by dust extinction for
$0.1\,\mu{m}$ Graphite and Silicate grains respectively. Thick (thin) lines correspond to
high (low) SFH. The errorbars are the errors on the duality test as expected
from upcoming SN Ia and BAO surveys. At $z<0.5$ the errors on angular diameter distance
measurements are not enough accurate for testing the duality.}\label{fig8}
\end{figure}

In figure~\ref{fig8} we plot Eq.~(\ref{dy}) 
for dust models A and B in the case of high (thick lines)
and low SFH (thin lines). The errorbars correspond to the expected
uncertainty of the distance-duality test. 
It can be seen that for high SFH, Silicate and Graphite particles of size $0.1\,\mu{m}$
would caused a clearly detectable violation of
the distance-duality relation.

\section{Conclusion}\label{conclu}

The goal of the next generation of SN Ia experiments is to determine the dark energy parameters
with high accuracy. For this to be possible systematic effects must be carefully taken
into account. Here we have studied the impact of intergalactic gray dust extinction. 
We have used an astrophysical motivated modeling of the IGM dust in terms of
the star formation history of the Universe and the physical properties of the dust grains.
We have identified a number of models which satisfy current astrophysical constraints
such those inferred from X-ray quasar halo scattering and the amplitude of the FIRB emission.
Although characterized by negligible reddening IGM dust may cause large extinction effects
and strongly affect the dark energy parameter estimation. In particular for high star-formation
history we find that dust particles with size $\sim0.1\,\mu{m}$ and a total dust density 
$\Omega_{\rm dust}^{\rm IGM}\sim10^{-6}$ may bias the inferred values of a constant dark energy
equation of state up $20\%$. 
Current SN Ia data are insensitive to such effects since the amplitude of the induced extinction is well
within the supernova magnitude dispersion.
Near-IR color analysis would require an accuracy
better than $1\%$ to detect the signature of these IGM dust particles. On the other hand
IGM dust arising from high SFH can be distinguished from the decrement of Balmer lines with 
high signal-to-noise spectroscopy.
We have also shown that cosmic dust violates the distance-duality relation, and depending on the dust model
this may be detected with future SN Ia and baryon acoustic oscillation data.

It is worth remarking that a number of caveats concerning the physics of the IGM
have been assumed throughout this analysis. Specifically we have considered 
a redshift independent dust-to-total-metal mass ratio. Unfortunately 
we still lack of a satisfactory understanding of the intergalactic medium both theoretically
and observationally which would allow us to make more robust prediction about IGM dust extinction. 
Indeed if we happen to live in a Universe with a total gray dust density 
$\Omega_{\rm dust}^{\rm IGM}\sim10^{-6}$ extinction effects on SN Ia observations
must be considered more than previously thought. 
The risk is to miss the discovery of the real nature of dark energy.

\section*{Acknowledgments}
It is a great pleasure to thank Bruce Bassett, Ed Copeland, Arlin Crotts, Zoltan Haiman, 
Dragan Huterer, Martin Kunz, Eric Linder, Frits Paerels and Raffaella Schneider for 
their helpful comments and suggestions. 
The author is supported by Columbia Academic Quality Fund.

\end{document}